\documentclass[sigconf]{acmart}
\AtBeginDocument{%
  \providecommand\BibTeX{{%
    \normalfont B\kern-0.5em{\scshape i\kern-0.25em b}\kern-0.8em\TeX}}}

\usepackage{multirow}
\usepackage{comment}
\usepackage{lipsum}
 
\newcommand\blfootnote[1]{%
  \begingroup
  \renewcommand\thefootnote{*}\footnotetext{#1}%
  \addtocounter{footnote}{-1}%
  \endgroup
}

\newtheorem{definition}{Definition}

\newtheorem{property}{Property}

\newcommand{\tabincell}[2]{\begin{tabular}{@{}#1@{}}#2\end{tabular}}  

\begin{document}

\title{SetRank: Learning a Permutation-Invariant Ranking Model for Information Retrieval}


\author{Liang Pang${}^{1}$, Jun Xu${}^{2,3,*}$, Qingyao Ai${}^{4}$, Yanyan Lan${}^{1}$, Xueqi Cheng${}^{1}$, Jirong Wen${}^{2,3}$ }
\affiliation{%
   \institution{${}^{1}$CAS Key Lab of Network Data Science and Technology, \\ Institute of Computing Technology, Chinese Academy of Sciences, Beijing, China}
}
\affiliation{%
	\institution{${}^{2}$Gaoling School of Artificial Intelligence, Renmin University of China\\${}^{3}$Beijing Key Laboratory of Big Data Management and Analysis Methods \;\;
	${}^{4}$The University of Utah, USA}
}
\email{{pangliang, lanyanyan, cxq}@ict.ac.cn,
	   {junxu, jrwen}@ruc.edu.cn, aiqy@cs.utah.edu}


\copyrightyear{2020}
\acmYear{2020}
\setcopyright{acmcopyright}
\acmConference[SIGIR '20]{2020 ACM SIGIR International ACM SIGIR Conference on Research and Development in Information Retrieval}{July 25--30, 2020}{Xi'an, China}
\acmBooktitle{The 43rd ACM SIGIR International ACM SIGIR Conference on Research and Development in Information Retrieval (SIGIR '20), July 25--30, 2020, Xi'an, China}
\acmPrice{}
\acmDOI{}
\acmISBN{}

\begin{abstract}\blfootnote{Corresponding author}
In learning-to-rank for information retrieval, a ranking model is automatically learned from the data and then utilized to rank the sets of retrieved documents. Therefore, an ideal ranking model would be a mapping from a document set to a permutation on the set, and should satisfy two critical requirements: (1)~it should have the ability to model cross-document interactions so as to capture local context information in a query; (2)~it should be permutation-invariant, which means that any permutation of the inputted documents would not change the output ranking.
Previous studies on learning-to-rank either design uni-variate scoring functions that score each document separately, and thus failed to model the cross-document interactions; or construct multivariate scoring functions that score documents sequentially, which inevitably sacrifice the permutation invariance requirement.
In this paper, we propose a neural learning-to-rank model called \mbox{SetRank} which directly learns a \emph{permutation-invariant ranking model defined on document sets of any size}. 
\mbox{SetRank} employs a stack of (induced) multi-head self attention blocks as its key component for learning the embeddings for all of the retrieved documents jointly. The self-attention mechanism not only helps \mbox{SetRank} to capture the local context information from cross-document interactions, but also to learn permutation-equivariant representations for the inputted documents, which therefore achieving a permutation-invariant ranking model. 
Experimental results on three benchmarks showed that the \mbox{SetRank} significantly outperformed the baselines include the traditional learning-to-rank models and state-of-the-art Neural IR models.    

\end{abstract}

\maketitle

\section{Introduction} \label{sec:introduction}

Learning-to-rank has been extensively studied in both academia and industry. 
Usually, the task of learning-to-rank can be described as two steps.
First, the training step, where a ranking model that projects each query-document pair to a ranking score is constructed and learned from labeled data such as user clicks and relevance annotations. 
And second, the testing step, where the learned ranking model is applied to a set of documents retrieved for a new query and finally returns a ranked document list to the users. 

Traditional learning-to-rank models are usually designed on the basis of the {probability ranking principle} (PRP)~\cite{robertson1977probability}.
PRP assumes that each document has a unique probability to satisfy a particular information need.
Therefore, the ranking scores of documents are assigned separately and are independent to each other.
Despite widely adopted, the power of PRP-based learning-to-rank methods, however, have been proved to be limited~\cite{ai2018learning,jiang2018beyond,ai2019learning}.
First, independent scoring paradigms prevent traditional learning-to-rank models from modeling cross-document interactions and capturing local context information.  
As shown by previous studies on pseudo relevance feedback~\cite{lavrenko2017relevance} and query-dependent learning-to-rank~\cite{can2014incorporating}, incorporating local context information such as query-level document feature distributions can significantly improve the effectiveness of modern ranking systems.
Second, as pointed by Robertson~\cite{robertson1977probability}, PRP works document-by-document while the results of ranking should be evaluated request-by-request. 
Behavior analysis on search engine users manifest that user's interactions with information retrieval systems show strong \textit{comparison} patterns~\cite{joachims2005accurately,yilmaz2014relevance}.
In practice, search engine users often compare multiple documents on a result page before generating a click action.
Also, studies on query-document relevance annotations show that information from other documents in the same ranked list could affect an annotator's decision on the current document~\cite{scholer2011quantifying,yang2017relevance}, which challenge the basic hypothesis that relevance should be modeled independently on each document for a single information request.

Recently, a new group of learning-to-rank methods whose scoring functions take multiple documents as input and jointly predict their ranking scores have received more and more attention.
For example, Ai et al.~\cite{ai2018learning,ai2019learning} propose to learn context-aware query-level ranking model that takes a sequence of documents as its input.
By modeling and comparing multiple documents together, a new scoring functions, namely the \textit{multivariate scoring functions}, naturally capture the local context information and produce the state-of-the-art performance on many learning-to-rank benchmarks.
Nonetheless, existing multivariate scoring approaches is sensitive to the order
of the inputted documents, which violates  the \textit{permutation invariance} requirement for ranking model.
That is because the models used by previous works, e.g. dense neural network~\cite{ai2019learning} and recurrent neural network~\cite{ai2018learning}, all assume that the input is a sequence of documents and heavily biased to its initial ranking. When the initial ranking has poor performance or disturbed by accident, the performances of these models decline significantly.  

Inspired by the work of Set Transformer~\citep{Lee:ICML19:SetTransformer}, in this paper, we propose to develop a multivariate ranking model whose input is a document set of any size, and the output is a permutation over the set. 
The model, referred to as SetRank, is to learn a multivariate scoring function mapping from document set to its permutation, making full use of self-attention mechanism~\cite{vaswani2017attention}. Firstly, It receives a whole set of documents (encoded as a set of vectors) as input. Then, employs a stack of (induced) multi-head self attention blocks as its key component for learning the embeddings for the whole set of documents jointly. Finally, the output of the last block is considered as the representation of the documents, and a row-wise fully connected network is used to generate the ranking scores.

SetRank offers several advantages. 
First of all, similar to existing work on multivariate scoring functions, SetRank considers the inputted documents as a whole via self-attention mechanism and model the interrelationship between them in ranking, which makes them more competitive than traditional learning-to-rank methods based on univariate scoring functions.
Secondly, as shown in the theoretical analysis, SetRank learns a permutation-invariant function as its ranking model, that is, any permutation of the inputted documents does not change the outputted document ranking. 
Additionally, self-attention is an additive function which is insensitive to the size of document set, while a sequential model, such as recurrent neural network, using nested function is severely affected by the size of document set. 
Compared to existing learning-to-rank models which learn multivariate ranking functions defined on predefined document sequences, SetRank is more natural for document ranking and more friendly for parallel computing. 
Last but not least, with the help of ordinal embeddings, SetRank is enabled to involve multiple document rankings as the initial rankings.

Experimental results on three large scale publicly available benchmarks showed that SetRank significantly outperformed the baselines including the traditional learning-to-rank models, such as RankSVM~\cite{joachims2006training}, RankBoost~\cite{freund2003efficient}, Mart~\cite{friedman2001greedy}, and LambdaMART~\cite{burges2010ranknet}, and state-of-the-art neural IR models, such as DLCM~\cite{ai2018learning} and GSF~\cite{ai2019learning}. Analyses showed that the ranker learned by SetRank is robust to the input orders and sizes, making it to be a stable ranking model. 

\section{Related Work} \label{sec:related_work}

This section reviews previous studies 
on learning-to-rank, 
deep learning for ranking, and the machine learning defined on sets. 

\subsection{Learning-to-Rank}
learning-to-rank refers to a group of algorithms that apply machine learning techniques to solve ranking problems.
It usually represents each query-document pair with a feature vector created by human or extracted with some data pre-processing techniques.
Then, the feature vectors are fed into a machine learning model to produce scores for each document so that the final ranking can be generated by sorting documents according to their scores.

In the past two decades, learning-to-rank methods based on different machine learning algorithms~\cite{friedman2001greedy, burges2005learning, joachims2006training} has been proposed and applied to a variety of ranking tasks including document retrieval~\cite{liu2009learning}, question answering~\cite{yang2016beyond}, recommendation~\cite{duan2010empirical}, conversational search~\cite{zhang2018towards}, etc.
Based on their loss functions, existing learning-to-rank algorithms can be categorized into three groups: pointwise~\cite{friedman2001greedy}, pairwise~\cite{joachims2006training,burges2005learning}, and listwise~\cite{cao2007learning,xia2008listwise,Bruch2019,Bruch2020}.
Despite their differences on loss functions, most traditional learning-to-rank methods are constructed based on the probability ranking principle~\cite{robertson1977probability}, which assumes that documents should be ranked individually according to their probabilities of satisfying a particular information need.
Under this assumption, they compute the ranking scores of each document separately based on their feature vectors, which is referred to as the uni-variate scoring paradigm.
Through intuitive, the uni-variate scoring paradigm is sub-optimal because it limits the model's power on capturing inter-document relationships and local context information, which is important for understanding user relevance in online systems~\cite{ai2019learning}.

To solve this problem, previous studies propose a new ranking paradigm that directly rank a list of documents together with a multivariate scoring function.
For example, \citet{ai2018learning} propose to encode the local context of an initial rank list with a recurrent neural network and rerank the documents based on the latent context embedding of all documents.
\citet{bello2018seq2slate, jiang2018beyond} propose a slate optimization framework that directly predict the ranking of a list of documents by jointly considering their features together. 
\citet{ai2019learning} further formalize the idea and propose a multivariate scoring framework based on deep neural networks. 
\cite{pasumarthi2019selfattentive} uses self-attention network to aggregate cross-documents information.
In recommendation, \citet{Pei:2019:PRR:3298689.3347000} propose to construct a list-wise re-ranking framework that personalizes recommendation results by jointly considering
multiple items in the same ranked list.

In this work, we propose a new learning-to-rank model based on the multivariate scoring paradigm. 
Different from previous studies that conduct reranking based on an initial order of documents~\cite{ai2018learning,bello2018seq2slate,ai2019learning}, our method is permutation-invariant and can be applied directly to a set of documents with or without any preprocessing.

\subsection{Deep Learning for Ranking}
Recently, deep learning methods such as deep neural networks and convolution neural networks have been widely applied to IR problems.
They have been proved to be effective in capturing latent semantics and extracting effective features for ranking problems.
For example, \citet{huang2013learning} propose a deep structured semantic model for ad-hoc retrieval that directly predict the relevance of each document based on a multi-layer feed-forward network and the embedding of trigrams in document titles. \citet{guo2016deep} analyze the key factors of relevance modeling and propose a deep relevance matching model that ranks documents based on the matching histogram of each query and document in the latent semantic space.
Later, \citet{pang2017deeprank} propose a DeepRank model that jointly considers the strength and patterns of query-document matching and uses a recurrent neural network to measure and aggregate the local relevance information for ranking.
Another type of deep models, originated from NLP, treat retrieval problems as a text matching between queries and documents. 
For example, \citet{hu2014convolutional} propose a group of ARC models that build text representations based on word embeddings and convolution neural networks for matching proposes. \citet{pang2016text} propose to match text based on a word matching matrix and propose a MatchPyramid model for paraphrase identification and question answering. \citet{wan2016match} further propose a Match-SRNN model that uses a spatial recurrent neural network to capture the recursive matching structure in text matching problems.

The proposed model in this paper can be considered as a deep model for IR because we take the self-attention networks~\cite{vaswani2017attention}, a popular neural technique used in machine learning tasks~\cite{devlin-etal-2019-bert,zhang2019self}, as our building blocks.
Different from previous models that directly extract features from raw text, we focus on a general task of learning-to-rank and assume that feature representation for each query-document pair has been computed in advance.
However, it's worth noting that our model framework also supports the joint learning of features and ranking models.
Also, in spite of their structures, most existing deep learning models for IR is confined to the same uni-variate scoring paradigm used by traditional learning-to-rank algorithms, while our model is a listwise ranking model under the multivariate scoring paradigm.

\subsection{Machine Learning Defined on Sets}
Typical machine learning algorithms such as regression or classification are designed for processing fixed dimensional data instances. Recent studies showed that they can be extended to handle the case when the inputs are permutation-invariant sets~\cite{zaheer2017deep,Edwards:ICLR17:NeuStat,Lee:ICML19:SetTransformer}. For example,~\citet{Edwards:ICLR17:NeuStat} demonstrated neural statistician which is an extension of variational autoencoder and can learn statistics of datasets. \citet{zaheer2017deep} theoretically characterize the permutation-invariant functions and provide a family of functions to which any permutation-invariant objective function must belong. \citet{Lee:ICML19:SetTransformer} further present the Set Transformer model to capture the interactions among elements in the input set. In this paper, we borrow the self-attention mechanism proposed in Transformer to capture the interactions among documents in ranking. 

\section{Ranking Model Defined on Sets} \label{sec:motivation}
In this section, we formalize document ranking as learning a multivariate scoring function defined on document sets, and then analyze the requirements of the function.

\subsection{Problem Formulation} \label{sec:formulation}
Generally speaking, the task of ranking aims to measure the relative order among a set of items under specific constrains. When ranking is completed, the arranged items is returned, called a permutation. 
When being applied to learning-to-rank for information retrieval, the constraint corresponds to a query $q$, and the set of items correspond to a set of $N$ documents $D = [d_i]_{i=1}^N \subseteq \mathbb{D}$ retrieved by the query $q$, and $\mathbb{D}$ is the set of all indexed documents. 
The task of ranking is to find a permutation $\pi\in \Pi_N$ on the document set $D$ so that some utility is maximized, where $\Pi_N$ is the set of all permutations of indices $\{1, 2, \cdots, N\}$.

In learning-to-rank, the ranking models are trained with a set of labeled query-document pairs. For each query $q$, its retrieved documents and the corresponding relevance labels are provided and denoted as $\psi_q = \{D = \{{d}_i\}, \mathbf{y}=\{y_i\} | 1 \leq i \leq N \}$, where $y_i$ denotes the relevance label corresponds to ${d}_i$. Therefore, the dataset that contains all of the training queries can be denoted as $\Psi=\{\psi_q\}$. The goal of learning-to-rank is to minimize the empirical loss over the training data and to yield the optimal scoring function $F(\cdot)$:
\begin{equation}
	\mathcal{L}(F) = \frac{1}{|\Psi|} \sum\nolimits_{\{D, \mathbf{y}\} \in \Psi}{l(\mathbf{y}, F(D))},
\end{equation}
where $l(\cdot)$ is the loss function; $F(\cdot) : \mathbb{D}^N \mapsto \mathbb{R}^N$ is the scoring function which is responsible for assigning scores to the documents, so that the result permutation on $D$ can be generated by sorting with the scores: $\hat{\pi}= sort \circ F(D)$. The operator `$sort$' sorts the documents in $D$ according to the scores assigned by $F(D)$ in descending order. 

Most of the ranking models learnt by the existing learning-to-rank models (including point-wise, pair-wise, and list-wise models) are uni-variate scoring functions $g:\mathbb{D} \mapsto \mathbb{R}$. They are special cases of $F$ because $F$ can be written as
\begin{equation}\label{eq:pointwiseScore}
	F(D)|_i = g(d_i), 1 \le i \le N,
\end{equation}
where $F(D)|_i$ is the $i$-th dimension of $F(D)$.
It is a natural and direct choice that fits the probability ranking principle~\cite{robertson1977probability}. 

\subsection{\mbox{Requirements of Ranking Model}} \label{sec:requirements}

We now show that a scoring function $F$ defined on document sets should satisfy two critical requirements. First, it should be able to characterize the inter-relationship among the retrieved documents, called cross-document interaction.  
Second, the output of the function should not be changed under any permutation of the documents in the input set, called permutation invariance. 

\subsubsection{Cross-document Interactions}
Although the PRP principle~\cite{robertson1977probability} and the derived uni-variate scoring functions have been widely used in information retrieval, recent studies show that they have several limitations in real settings. The major assumption in PRP is that the relevance of a document to a query is independent of the relevance of other documents. However, \citet{fuhr2008probability} pointed out that ``the relevance of any additional relevant document clearly depends on the relevant documents seen before''; \citet{robertson1977probability} also pointed out that ``PRP works document-by-document while the results of ranking should be evaluated request-by-request''; \citet{scholer2011quantifying,yang2017relevance} studied the relevance annotations and showed that information from other documents in the same ranked list could affect an annotator's decision on the current document. 

Therefore, the assumptions underlying the classical PRP and the uni-variate scoring function shown in Equation~(\ref{eq:pointwiseScore}), is hard to hold in some settings. It is expected that the multivariate scoring function $F$ can deal with a more general ranking problem of taking account of the dependency information and cross-document interactions. 

\subsubsection{Permutation Invariance}
The input to a ranking model is a set that contains multiple documents. According to the definition of a set, the response of a function defined on a set should be ``indifferent'' to the ordering of the elements. That is, the ranking model should satisfy the permutation invariance property: the target ranking for a given set is the same regardless of the order of documents in the set. 
That is, if we exchange the input positions of $d_i$ and $d_j$, it will not affect the result of the final ranking. 
In practice, input sequence often convey covert bias information, e.g. order by document IDs, order by create time, or order by time-consuming, which may not correlate with the usefulness of the documents and could reduce the robustness of permutation awareness model.
A permutation-invariant model addresses this problem with theoretical guarantees.

One way to construct a permutation-invariant ranking model is first assigning relevance scores to the documents with a permutation equivariant scoring function, and then sorting according to the scores. \citet{Lee:ICML19:SetTransformer} give a definition of permutation equivariant scoring function, as shown in Definition~\ref{def:PermEqu}. 
\begin{definition}\cite{Lee:ICML19:SetTransformer}\label{def:PermEqu}
 Let $\Pi_N$ be the set of all permutations of indices $\{1, \cdots, N\}$. A function $F : X^N \to Y^N$ is permutation equivariant if and only if $\forall \pi \in \Pi_N$, 
 \[
 F([d_{\pi(1)}, \cdots, d_{\pi(N)}]) = [F(D)|_{\pi(1)}, \cdots, F(D)|_{\pi(N)}],
 \] 
 where $D = [d_1, \cdots, d_N]$ is a set with the object order $1, 2, \cdots, N$, and $F(D)|_{\pi(i)}$ is the $\pi(i)$-th dimension of $F(D)$. 
\end{definition}

An example of a permutation equivariant function in IR ranking is the uni-variate scoring function shown in Equation~(\ref{eq:pointwiseScore}). 

\section{Our Approach: SetRank} \label{sec:approach}
In this section, we penetrate into the details of our proposed SetRank model, and introduce the training method.

\subsection{Overall Architecture} \label{sec:architecture}
\begin{figure}
    \centering
    \includegraphics[width=0.85\linewidth]{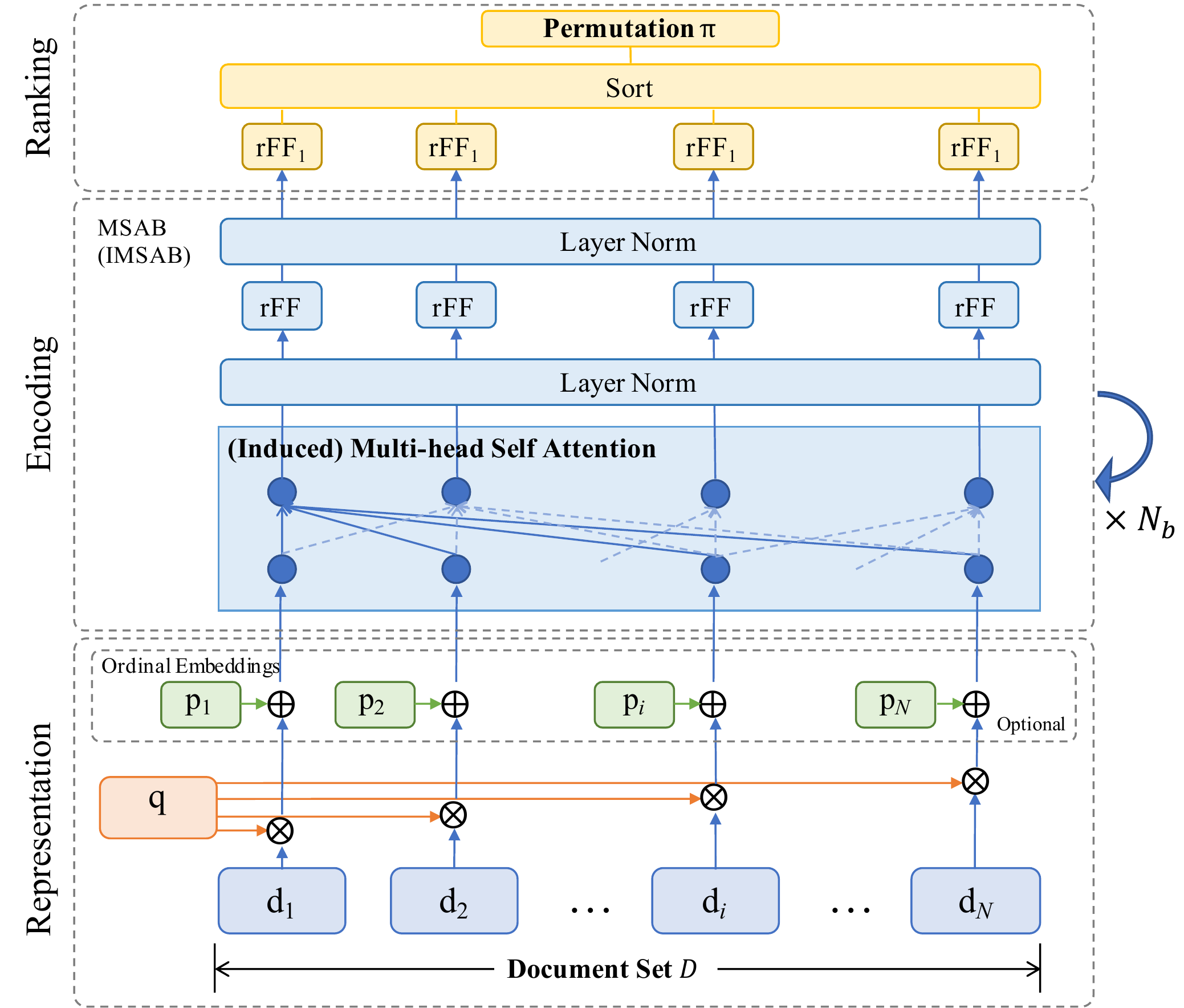}
    \caption{Architecture of SetRank. The representation layer generates representation of query-document pairs separately; the encoding layer jointly process the documents with multiple sub-layers of MSAB (or IMSAB), and output internal document representations; the ranking layer calculates the scores and sorts the documents.}
    \label{fig:architecture}
\end{figure}

Inspired by the Transformer for machine translation~\cite{vaswani2017attention,dehghani2018universal} and language modeling~\cite{devlin-etal-2019-bert}, we have devised a novel ranking model defined on document sets and satisfy the above requirements. The model is referred to as `SetRank' and is shown in Figure~\ref{fig:architecture}. SetRank makes use of multi-head self attention blocks and its modifications to construct a multivariate scoring function that satisfies the {\em cross-document interactions} and the {\em permutation invariance} requirements, additionally adapts to the set input of any size ({\em set size adaptation}).

The pipeline of document ranking in SetRank consists of three layers: representation, encoding, and ranking. 
First, the representation layer separately represents each inputted document as a vector of features such as the hand-crafted features used in traditional learning-to-rank models.
Also, it can involve the initial ranking of the documents in feature representations through the ordinal embeddings. 
The initial rankings could be generated by existing ranking models such as BM25 or a trained LambdaMART model.
Second, the encoding layer enriches each query-document pair feature vector by involving other feature vectors of the associated documents. In this paper, we use Multi-head Self Attention Block (MSAB) or Induced Multi-head Self Attention Block (IMSAB) to take a set of query-document pairs representations as their input and generate a set of new representations using the self-attention blocks. Multiple sub-layers of MSAB or IMSAB blocks are stacked together with identical structure for modeling the high-order interactions between the documents. Third, the ranking layer receives the output vectors of the top-most MSAB (or IMSAB) block, passes them to a row-wise feed-forward (rFF) function, generates relevance scores for all of the documents, and finally sorts the documents according to these scores.
The following sections will introduce the details of the three layers. 

\subsection{Document Representation} \label{sec:representation}
Given a query $q$ and its associated document set $D=[d_1,d_2, \cdots, d_N]$, each of the document in $D$ can be represented as a feature vector 
\[
\mathbf{d}_i = \phi(q, d_i),\; \mathrm{where} \; \mathbf{d}_i \in \mathbb{R}^{E},
\] 
where $\phi$ is the function for feature extraction and $E$ is the dimension of the vector. The features extracted in traditional learning-to-rank are used here, including document only feature of PageRank, query-document matching features of TF-IDF, BM25 etc. In the experiments of this paper, we used the features provided by the benchmark datasets. 

Besides the traditional learning-to-rank features, SetRank can optionally include the document ordinal embeddings as the input. In real search engines, the associated documents may have some prior rankings generated by the default ranking models such as BM25 or LambdaMART etc. To involve these initial ranking information and inspired by the positional embedding in Transformer~\cite{vaswani2017attention}, we propose an ordinal embedding function $P$ which takes the absolute ranking position of a document as input, and encodes the position to a vector as the same dimension with $\mathbf{d}_i$:
\[
\mathbf{p}_i = P(\mathrm{rank}(d_i)),\; \mathrm{where} \; \mathbf{p}_i \in \mathbb{R}^{E},
\]
where $\mathrm{rank}(d_i)$ denotes the absolute rank position of $d_i$ in the initial ranking generated by models such as BM25, LambdaMART etc. 

The ordinal embedding vectors and the learning-to-rank feature vectors are respectively added, forming a feature matrix $\mathbf{X}\in \mathbb{R}^{N \times E}$ for the $N$ retrieved documents: 
\[
\mathbf{X} = [\mathbf{d}_1 + \mathbf{p}_1, \mathbf{d}_2 + \mathbf{p}_2,\cdots, \mathbf{d}_N + \mathbf{p}_N]^T.
\]

Note that in some cases we may have more than one initial rankings because different ranking model (e.g., BM25 and LM4IR) may be applied simultaneously. The ordinal embeddings of a document corresponding to different rankings can be summed together, forming an overall ordinal embedding vector. In this way, SetRank is enabled to input multiple initial rankings. Also note that though the ordinal embedding mechanism is similar to the position embedding in Transformer, they are different in that the position embedding characterizes the position of a word in a sequence, while the ordinal embedding in SetRank characterizes the rank of a document w.r.t. a ranking algorithm being applied to a document set. Therefore, each word inputted to Transformer has only one position embedding while each document in SetRank may have multiple ordinal embeddings. 

\subsection{Document Encoding with (Induced) Multi-head Self Attention Block} \label{sec:encoding}

\begin{figure}
    \centering
    \includegraphics[width=0.85\linewidth]{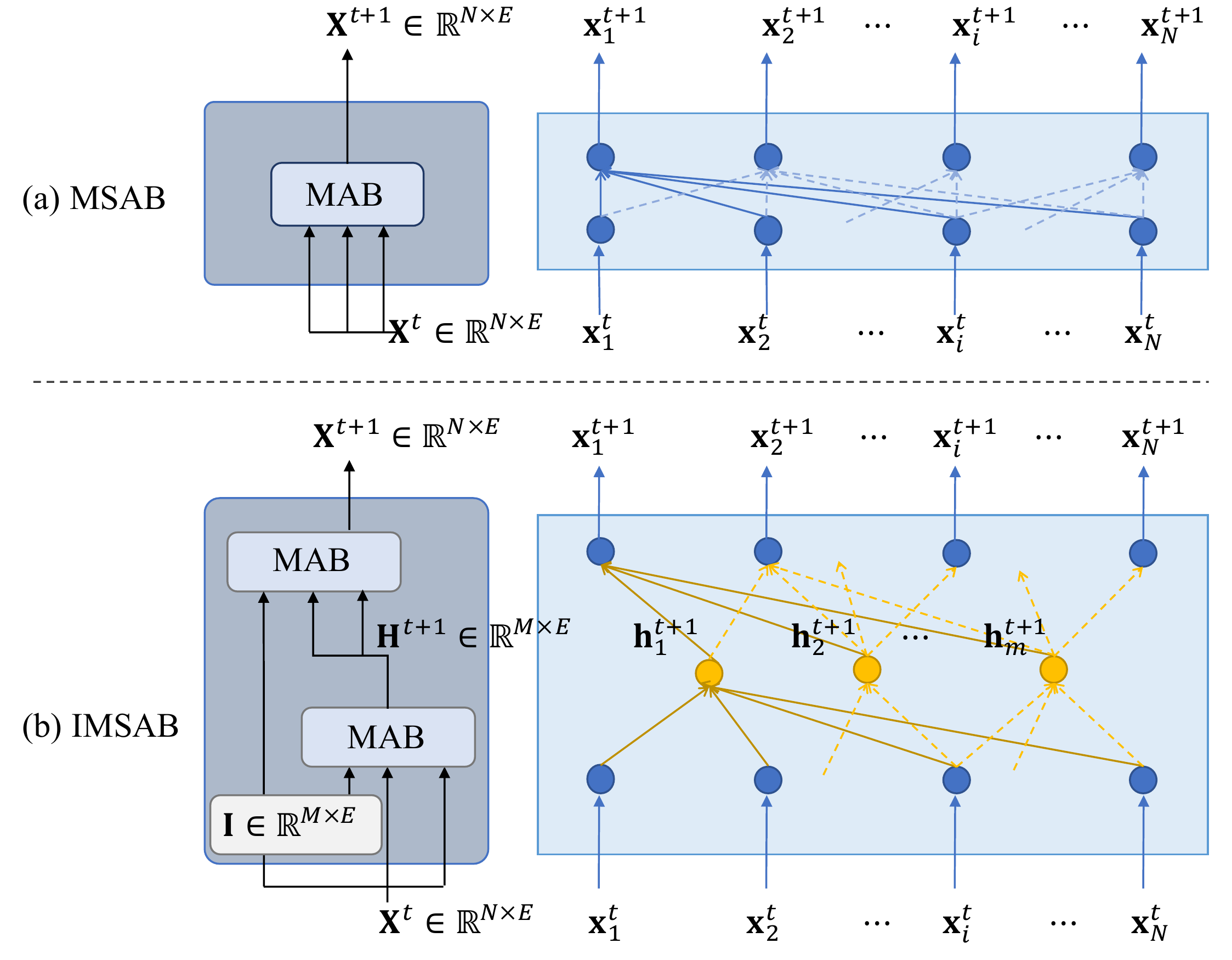}
    \caption{(a) The structure of multi-head self attention block (MSAB). MSAB encodes an arbitrary set $\mathbf{X}^t$ of size $N$ and output an set $\mathbf{X}^{t+1}$ of size $N$. (b) The structure of induced multi-head self attention block (IMSAB) with induced size of $M$. IMSAB first encodes an arbitrary set $\mathbf{X}^t$ of size $N$ to a fixed set $\mathbf{H}^{t+1}$ of size $M$, and then encodes $\mathbf{H}^{t+1}$ to an $N$-size output $\mathbf{X}^{t+1}$. IMSAB can be interpreted as first creating $M$ cluster centers and then encoding the inputted $N$ documents using these $M$ cluster centers.}
    \label{fig:block}
\end{figure}

The key of SetRank is the encoding component which takes the document representations $\mathbf{X}^{0} = \mathbf{X} \in {\mathbb{R}^{N \times E}}$ as input, and jointly encoding these documents as their internal codes $\mathbf{X}^{N_b}\in \mathbb{R}^{N \times E}$. 
The encoding component is a stack of $N_b$ multi-head self attention blocks (MSAB) (or induced multi-head self attention blocks (IMSAB)) with identical structures. Each MSAB (or IMSAB) receives a set of $N$ vectors (packed as a matrix $\mathbf{X}^{t}, 0 \le t \le N_b-1$), processes the vectors with the (induced) multi-head self attention layer followed by a layer normalization, and then through a row-wise feed-forward network (rFF) layer followed by another layer normalization. Finally, it sends the vectors $\mathbf{X}^{t+1}$ as output to the next layer of MSAB (or IMSAB). 

Next, we introduce the details of MSAB and IMSAB. 

\subsubsection{Multi-head self attention block (MSAB)} \label{subsec:MSAB}
MSAB is based on the attention mechanism in deep neural networks. As have shown in \cite{vaswani2017attention}, the attention function can be formalized as a scaled dot-product attention with three inputs: 
\begin{equation}\label{eq:attention}
	\mathrm{Attn}(\mathbf{Q}, \mathbf{K}, \mathbf{V}) = \mathrm{softmax}\left(\frac{\mathbf{QK}^T}{\sqrt{E}}\right)\mathbf{V}, 
\end{equation}
where $\mathbf{Q} \in \mathbb{R}^{N_q \times E}$ denotes the attention query matrix\footnote{Note that the ``query'' in the attention function and IR are with different meanings. }, $\mathbf{K} \in \mathbb{R}^{N_k \times E}$ the key matrix, and  $\mathbf{V} \in \mathbb{R}^{N_k \times E}$ the value matrix. $N_q, N_k$, and $E$ denote the number of attention query, keys/values vectors, the dimensions of the representation, respectively. The attention mechanism can be explained as: for each attention query vector in $\mathbf{Q}$, it first computes the dot products of the attention query with all keys, aiming to evaluate the similarity between the attention query and each key. Then, it is divided each by $\sqrt{E}$, and applies a softmax function to obtain the weights on the values. Finally, the new representation of the attention query vector is calculated as weighed sum of values.

To make the attention function more flexible, additional multi-head strategy is usually combined with the attention mechanism. That is, instead of computing a direct attention function, multi-head strategy first projects the inputs $\mathbf{Q}, \mathbf{K}, \mathbf{V}$ into $h$ different spaces, each has the dimensionality of $\hat{E} = E / h$ and projects with the attention function $Attn(\cdot, \cdot, \cdot)$ defined in Equation~(\ref{eq:attention}):
\begin{equation}\label{eq:Multihead}
	\mathrm{Multihead}(\mathbf{Q}, \mathbf{K}, \mathbf{V}) =
		\mathrm{concat}\left(\left[ \mathrm{Attn}(\mathbf{Q W}_i^Q, \mathbf{K W}_i^K, \mathbf{V W}_i^V) \right]_{i=1}^h \right)
\end{equation}
where $\mathbf{W}_i^Q, \mathbf{W}_i^K, \mathbf{W}_i^V \in \mathbb{R}^{E \times \hat{E}}$ reshape the input matrices, and the output $\mathrm{Multihead}(\mathbf{Q}, \mathbf{K}, \mathbf{V})$ has the same shape with $\mathbf{Q}$.

Based on the Multihead function in Equation (\ref{eq:Multihead}), the multi-head attention block (MAB) is defined as
\begin{equation}
	\begin{split}
		\textrm{MAB}(\mathbf{Q}, \mathbf{K}, \mathbf{V}) &= \textrm{LayerNorm}(\mathbf{B} + \textrm{rFF}(\mathbf{B})), \\
		\textrm{where} \; \mathbf{B} &= \textrm{LayerNorm}(\mathbf{Q} + \textrm{Multihead}(\mathbf{Q}, \mathbf{K}, \mathbf{V}))
	\end{split}
\end{equation}
where $\textrm{rFF}(\cdot)$ is a row-wise feedforward layer, and $\textrm{LayerNorm}(\cdot)$ is layer normalization~\cite{ba2016layer}. The MAB is an adaptation of the encoder block of the Transformer~\cite{vaswani2017attention} without positional encoding and dropout. Ideally, $\mathbf{K}$ and $\mathbf{V}$ can be treated as the index and context of a dictionary, while $\mathbf{Q}$ is the information need of the user. For simplicity, we set $\mathbf{K} = \mathbf{V}$, that means, the index is the context itself.

Finally, the self-attention in MSAB is a special case of MAB when  $\mathbf{Q} = \mathbf{K} = \mathbf{V}$, as shown in Figure~\ref{fig:block}(a). In MSAB, the model represents each document using all the document in the input set:
\begin{equation}
	\textrm{MSAB}(\mathbf{X}) = \textrm{MAB}(\mathbf{X}, \mathbf{X}, \mathbf{X}).
\end{equation}

\subsubsection{Induced Multi-head Self Attention Block (IMSAB)} \label{subsec:IMSAB}
One problem with MSAB is its sensitivity to the size of input set, as will be shown in our experimental studies. The MSAB trained on the data that each query has $N_0$ labeled documents does not work well on the test time if the queries are associated with $N_1> N_0$ documents. In most cases, $N_0$ is relatively small because of the expensive and time-consuming relevance label annotation. In real applications, however, it is not nature and realistic to limit the the number of documents that a query can retrieve. Previous studies showed that other multivariate scoring functions also have similar limitations~\cite{ai2019learning}.

To address the problem, we employ the induced multi-head self attention~\cite{Lee:ICML19:SetTransformer}, or IMSAB, for enabling the model to process the inputted document set with any size.  
As shown in Figure~\ref{fig:block}(b), IMSAB can be spilt into two steps. 
First, it constructs $M$ fake attention query vectors, denoted as $\mathbf{I} \in \mathbb{R}^{M \times E}$, to extract information from original keys/values $\mathbf{X} \in \mathbb{R}^{N \times E}$. Usually, $M < N$ and the $M$ vectors in $\mathbf{I}$ can be viewed as the index of cluster centers for keys/values. The result $\mathbf{H} \in \mathbb{R}^{M \times E}$ is the $M$ cluster centers for keys/values. 
Then, the original documents $\mathbf{X} \in \mathbb{R}^{N \times E}$ find their contextual representations based on the $M$ cluster centers $\mathbf{H} \in \mathbb{R}^{M \times E}$. The new representations generated by $\mathrm{IMSAB}_M(\mathbf{X})$ have the same shape with the original $\mathbf{X}$ and can be formally written as:
\begin{equation}\label{eq:IMSAB}
	\begin{split}
		\mathrm{IMSAB}_M(\mathbf{X}) =  \mathrm{MAB}(\mathbf{X}, \mathbf{H}, \mathbf{H}), \;
		\mathrm{where} \; \mathbf{H} = \mathrm{MAB}(\mathbf{I}, \mathbf{X}, \mathbf{X}).
	\end{split}
\end{equation}

\subsection{Document Ranking} \label{subsec:ranking}
Stacking aforementioned MSAB or IMSAB$_{M}$ blocks and passing the results to row-wise feedforward neural networks, we achieve two versions of SetRank scoring functions, respectively named as SetRank$_{\textrm{MSAB}}$ and SetRank$_{\textrm{IMSAB}}$:
\begin{equation}\label{eq:RankSetScore}
	\begin{split}
	    &\textbf{X}^{N_b}_{\mathrm{MSAB}} = \underbrace{\textrm{MSAB}(\textrm{MSAB} \dots (\textrm{MSAB}}_{N_b}(\mathbf{X^0}))), \\
		&\textrm{SetRank}_{\textrm{MSAB}}(D) = \textrm{rFF}_1 \left( \textbf{X}^{N_b}_{\mathrm{MSAB}}  \right), \\
	\end{split}
\end{equation}
where $\textrm{rFF}_1$ is a row-wise feed-forward network that projects each document representation into one real value as the corresponding ranking score. 
Similarly, $\textrm{SetRank}_{\textrm{IMSAB}}$ can be obtained by replacing MSAB to IMSAB. 
The final document rankings, thus, can be achieved by sorting the documents according to the scores:
\begin{equation}\label{eq:RankSetPermu}
\begin{split}
    \hat{\pi}_{\textrm{MSAB}} & = sort \circ \textrm{SetRank}_{\textrm{MSAB}}(D).
\end{split}
\end{equation}

\subsection{Model Training} \label{sec:training}
Following the practices in~\cite{ai2018learning}, we adopt an attention rank loss function for model training. The basic idea is to measure the distance between an attention distribution generated by the ranking scores and that of generated by the relevance judgments.
 
Given a labeled query $\psi_q=\{D=\{d_i\}, \mathbf{y}=\{y_i\}|1\leq i\leq N\}$, where $y_i$ is the relevance label and denotes the information gain of document $d_i$ for query $q$. The optimal attention allocation strategy in terms of relevance labels for document $d_i (i=1, \cdots, N)$ is:
\[
	a_i^y = \frac{\tau(y_i)}{\sum_{d_k \in D}\tau(y_k)}, \;\;\;\; \tau(x) = \left\{\
\begin{array}{cc}
	\exp(x)& x > 0\\
	0 & \textrm{otherwise} 
\end{array}
\right.
\] 

Similarly, given the predicted ranking scores $\{s_1, \cdots, s_N\}= \mathrm{SetRank}(D)$, the attention w.r.t. the predicted scores for document $d_i (i=1, \cdots, N)$ is:
\[
	a_i^s = \exp(s_i) \big/ \sum\nolimits_{d_k \in D}\exp(s_k).
\] 
Therefore, the list-wise cross entropy loss function can be constructed on the basis of these two attention distributions: 
\[
	\mathcal{L} = \sum\nolimits_{d_i \in D} a_i^y \log a_i^s + (1-a_i^y)\log(1-a_i^s).
\]

Note that SetRank is insensitive to the input set size, especially for SetRank based on IMSAB. In the document representation step, ordinal embeddings may be involved to utilize the initial rankings. However, it stop the model from working on the input set size larger than the largest set size in the training data, because the larger ordinal embeddings are unavailable. To address the issue, we proposed a relative ordinal embedding sampling strategy during the model training.
First, a max set size $N_{max}$ (could be a very large number) is chosen and the length of ordinal embeddings is also set to $N_{max}$. Then, for each training query, we sample the start number $s$, constructing a sequence of index numbers $[s, s+1, \cdots, s+N-1]$ are used to replace the original document ranking indexes $[1, 2, \cdots, N]$. With this trick, all of positions less than $N_{max}$ can be trained and have their ordinal embeddings. Finally, during the test phase, the ordinal embedding index larger than $N$ and less than $N_{max}$ is available, so that we can process ranking list smaller than $N_{max}$.


\subsection{Theoretical Analysis} \label{sec:theoretical}
The existing learning-to-rank algorithms either use uni-variate ranking function that assigns relevance scores to the retrieved documents independently, or use multivariate ranking functions defined on document sequence in which the output is sensitive to the input order of the documents. In contrast, SetRank tries to learn a permutation-invariant ranking model defined on document sets of any size. We show that the RankSet scoring functions defined in Equation~(\ref{eq:RankSetScore}) are permutation equivariant, proved in Section~\ref{appendix:proof}.
\begin{proposition}\label{pro:PermuEqu}
The multivariate scoring functions SetRank$_{\textrm{MSAB}}$ and SetRank$_{\textrm{IMSAB}}$ are permutation equivariant.	
\end{proposition}

\begin{property}
The multi-head attention block is permutation equivariant.	
\end{property}

\begin{property}
The induced multi-head self attention block is permutation equivariant.	
\end{property}
The property can be proved in a similar way. Also, \citet{zaheer2017deep} showed that the composition of permutation equivariant functions is also permutation equivariant. Thus, it is easy to show that SetRank$_{\textrm{MSAB}}$ and SetRank$_{\textrm{IMSAB}}$ are permutation equivariant.

Since the final outputted document ranking is achieved by sorting with the scores, it is obvious that the two variations of SetRank model shown in Equation~(\ref{eq:RankSetPermu}) are permutation-invariant.

\section{Experiments} \label{sec:experiments}
In this section, we describe our experimental settings and results on three publicly available large scale benchmarks.

\subsection{Experiment Settings} \label{sec:settings}

\subsubsection{Datasets} \label{subsec:datasets}
The experiments were conducted on three public available learning-to-rank datasets: Istella LETOR dataset (Istella)\footnote{\url{http://blog.istella.it/istella-learning-to-rank-dataset/}}, Microsoft LETOR 30K (MSLR30K)\footnote{\url{http://research.microsoft.com/en-us/projects/mslr/}} , and Yahoo! LETOR challenge set1 (Yahoo!)\footnote{\url{http://learningtorankchallenge.yahoo.com}}.
All these three datasets contain queries and documents sampled from real search engines. 
Each query-document pair was labeled by human annotators with 5-level relevance judgments ranging from 0 (irrelevant) to 4 (perfectly relevant).
As for features, MSLR30K represents each query-document pair with ranking features extracted by human exports; Yahoo! and Istella represent each query-document pair with the actually feature vectors used in the online systems.
In the experiments, we conducted 5-fold cross validation on MSLR30K, and used the predefined train/valid/test split in Yahoo! and Istella. 
More characteristics of the three datasets are listed in Table~\ref{Table.Datasets}.

\begin{table}
    \centering
    \caption{Dataset statistics.}
	\label{Table.Datasets}
	\small
	\begin{tabular}{l l l l}
		\hline
		Property & Istella & MSLR 30K & Yahoo!\\
		\hline
		\#features & 220 & 136 & 700 \\
		\#queries in training & 20,317 & 18,919 & 19,944 \\
		\#queries in validation & 2,902 & 6,306 & 2,994 \\
		Total \#docs in train & 7,325,625 & 2,270,296 & 473,134 \\
		\#queries in test & 9,799 & 6,306 & 6,983 \\
		Total \#docs in test & 3,129,004 & 753,611 & 165,660 \\
		Avg. \#docs per query in test & 319.31 & 119.51 & 23.72 \\ 
		\hline
	\end{tabular}
\end{table}

\subsubsection{Baselines and Evaluation Measures} \label{subsec:baselines}

Learning-to-rank models based on the uni-variate scoring functions were used as the baselines in the experiments. 
\textbf{\textsc{RankSVM}}~\cite{joachims2006training}: A classic pairwise learning-to-rank model built on the basis of SVM. 	
\textbf{\textsc{RankBoost}}~\cite{freund2003efficient}: Another classic pairwise learning-to-rank model based on boosting;
\textbf{\textsc{MART}}~\cite{friedman2001greedy}: A pointwise learning-to-rank algorithm based on gradient boosting regression trees;
\textbf{\textsc{LambdaMART}}~\cite{burges2010ranknet}: An extension of MART that directly optimize NDCG in training. LambdaMART has been widely used in commercial search engines.
The state-of-the-art deep learning-to-rank models based on multivariate scoring functions were also adopted as the baselines.
\textbf{\textsc{DLCM}}~\cite{ai2018learning}: A deep learning-to-rank model that use local context information to re-rank a list of documents. In DLCM, RNN is used to encode the initial document list. 
\textbf{\textsc{GSF}}~\cite{ai2019learning}: A groupwise scoring model that ranks documents based on group comparisons.

For RankSVM, we used the implementations in  SVM$^{\mathrm{rank}}$~\footnote{\url{https://www.cs.cornell.edu/people/tj/svm_light/svm_rank.html}}; For RankBoost, Mart, and LambdaMART, we used the implementations in \textit{RankLib}~\footnote{\url{https://sourceforge.net/p/lemur/wiki/RankLib/}}. For DLCM and GSF, we used the implementations and settings reported in Github repositories: \textit{DLCM}~\footnote{\url{https://github.com/QingyaoAi/Deep-Listwise-Context-Model-for-Ranking-Refinement}} and \textit{TF-Ranking}~\footnote{\url{https://github.com/tensorflow/ranking/tree/master/tensorflow_ranking}}.

All models were tuned and selected based on their performances on the validation set in terms of NDCG@10. Following the practices in~\cite{ai2018learning}, we set the RankSVM parameter $C=200$; the training rounds of RankBoost as 1000. For tree-based models, the number of trees is set to 1000 and the leaf number 20.

Normalized Discounted Cumulative Gain (NDCG)~\cite{jarvelin2002cumulated} is used to evaluate the performances of the ranking models. To show the model performances at different positions, we reported the NDCG values at the ranks of 1, 3, 5, and 10. 

\subsubsection{SetRank Settings} \label{subsec:setrank_settings}
We tested four variations of the SetRank, 
\textbf{SetRank${}_{\mathrm{MSAB}}$/SetRank${}_{\mathrm{IMSAB}}$}: using stacked MSAB/IMSAB for documents encoding, without ordinal embeddings;
\textbf{SetRank${}_{\mathrm{MSAB}}^{init}$/ SetRank${}_{\mathrm{IMSAB}}^{init}$}: using stacked MSAB/IMSAB for documents encoding, and ordinal embeddings based on the initial document ranking generated by LambdaMART;
In all of the experiments, the SetRank models were configured to have 6 stacked blocks, each has 256 hidden units and 8 heads. In order to fit the hidden dimension of the MSAB or IMSAB, we further appended a $\mathrm{rFF}_{256}$ on $\phi(q, d_i)$ to reshape the dimension of $\mathbf{d}_i$ to 256. 
For SetRank${}_{\mathrm{IMSAB}}$ and SetRank${}_{\mathrm{IMSAB}}^{init}$, the induced dimension $M$ is set to 20.
The gradient descent method Adam~\cite{kingma2014adam} with learning rate 0.001 is used to optimize the objective.
Following the practices in \cite{ai2018learning}, in all of the experiments we retrieved top 40 documents per query using LambdaMart. The ranking models were trained on these retrieved documents and then used to re-rank the documents.
The initial document rankings using in DLCM, SetRank${}_{\mathrm{MSAB}}^{init}$ and SetRank${}_{\mathrm{IMSAB}}^{init}$ are generated by LambdaMART. For fair comparison, we also evaluated DLCM without the initial rankings, denotes as DLCM${}^{w/o \, init}$.

The source code and experiments of SetRank can be found at \textbf{\url{https://github.com/pl8787/SetRank}}. 

\begin{table}
    \centering
    \caption{Performance comparison of different models on Istella, MSLR30K and Yahoo datasets. Significant performance improvements (paired t-test with $\mathbf{p}$-value $\leq$ 0.05) over LambdaMart and DLCM are denoted as `+' and `$\dag$', respectively. Boldface indicates the best performed results.}
	\label{Table.Experiments}

	\begin{tabular}{l l l l l}
		\multicolumn{5}{c}{(a) Ranking accuracies on Istella LETOR dataset} \\
		\hline
		 & \multicolumn{4}{c}{NDCG} \\
		Model & @1 & @3 & @5 & @10 \\
		\hline
		\hline
		RankSVM & 0.5269 & 0.4867 & 0.5041 & 0.5529 \\
		RankBoost & 0.4457 & 0.3977 & 0.4097 & 0.4511 \\
		Mart & 0.6185 & 0.5633 & 0.5801 & 0.6285 \\
		LambdaMart & 0.6571 & 0.5982 & 0.6118 & 0.6591 \\
		\hline
		\multicolumn{5}{c}{Without initial rankings}\\
		DLCM${}^{w/o \, init}$ & 0.6272 & 0.5717 & 0.5848 & 0.6310 \\
		GSF & 0.6224 & 0.5796 & 0.5968 & 0.6508 \\
		SetRank${}_{\mathrm{MSAB}}$ & 0.6702${}^{+\dag}$ & 0.6150${}^{+\dag}$ & 0.6282${}^{+\dag}$ & 0.6766${}^{+\dag}$ \\ 
		SetRank${}_{\mathrm{IMSAB}}$ & 0.6733${}^{+\dag}$ & 0.6136${}^{+\dag}$ & 0.6278${}^{+\dag}$ & 0.6737${}^{+\dag}$ \\ 
		\hline
				\multicolumn{5}{c}{With initial rankings generated by LambdaMart}\\
		DLCM & 0.6558 & 0.6030${}^+$ & 0.6194${}^+$ & 0.6680${}^+$ \\ 
		SetRank${}_{\mathrm{MSAB}}^{init}$ & 0.6745${}^{+\dag}$ & 0.6201${}^{+\dag}$ & \textbf{0.6350}${}^{+\dag}$ & 0.6819${}^{+\dag}$\\ 
		SetRank${}_{\mathrm{IMSAB}}^{init}$ & \textbf{0.6760}${}^{+\dag}$ & \textbf{0.6202}${}^{+\dag}$ & 0.6345${}^{+\dag}$ & \textbf{0.6834}${}^{+\dag}$ \\ 
		\hline

		\\
		
		\multicolumn{5}{c}{(b) Ranking accuracies on Microsoft LETOR 30K dataset} \\
		\hline
		 & \multicolumn{4}{c}{NDCG} \\
		Model & @1 & @3 & @5 & @10 \\
		\hline
		\hline
		RankSVM &  0.3010 & 0.3180 & 0.3350 & 0.3650 \\
		RankBoost & 0.2788 & 0.2897 & 0.3043 & 0.3339 \\
		Mart & 0.4436 & 0.4344 & 0.4414 & 0.4633 \\
		LambdaMart & 0.4570 & 0.4420 & 0.4450 & 0.4640 \\
		\hline
		\multicolumn{5}{c}{Without initial rankings}\\
		DLCM${}^{w/o \, init}$ & 0.3985 & 0.3919 & 0.4001 & 0.4245 \\ 
		GSF & 0.4129 & 0.4073 & 0.4151 & 0.4374 \\ 
		SetRank${}_{\mathrm{MSAB}}$ & 0.4243 & 0.4116 & 0.4177 & 0.4403 \\ 
		SetRank${}_{\mathrm{IMSAB}}$ & 0.4290 & 0.4166 & 0.4220 & 0.4428 \\ 
		\hline
		
		\multicolumn{5}{c}{With initial rankings generated by LambdaMart}\\
		DLCM & \textbf{0.4630}${}^+$ & 0.4450${}^+$ & 0.4500${}^+$ & 0.4690${}^+$ \\ %
		SetRank${}_{\mathrm{MSAB}}^{init}$ & 0.4572 & 0.4452${}^+$ & 0.4499${}^+$ & 0.4692${}^+$ \\ 
		SetRank${}_{\mathrm{IMSAB}}^{init}$ & 0.4591${}^+$ & \textbf{0.4469}${}^{+\dag}$ & \textbf{0.4515}${}^+$ & \textbf{0.4696}${}^+$ \\ 
		\hline

		\\		
		\multicolumn{5}{c}{(c) Ranking accuracies on Yahoo! LETOR challenge set1 dataset} \\
		\hline
		 & \multicolumn{4}{c}{NDCG} \\
		Model & @1 & @3 & @5 & @10 \\
		\hline
		\hline
		RankSVM &  0.6370 & 0.6500 & 0.6740 & 0.7260 \\
		RankBoost & 0.6293 & 0.6409 & 0.6661 & 0.7159 \\
		Mart & 0.6830 & 0.6827 & \textbf{0.7034} & \textbf{0.7469} \\
		LambdaMart & 0.6770 & 0.6760 & 0.6960 & 0.7380 \\
		\hline
		\multicolumn{5}{c}{Without initial rankings}\\
		DLCM${}^{w/o \, init}$ & 0.6693 & 0.6751 & 0.6958 & 0.7391 \\
		GSF & 0.6429 & 0.6604 & 0.6838 & 0.7316 \\
		SetRank${}_{\mathrm{MSAB}}$ & 0.6623 & 0.6698 & 0.6911 & 0.7369 \\ 
		SetRank${}_{\mathrm{IMSAB}}$ & 0.6711 & 0.6760 & 0.6960 & 0.7398 \\ 
		\hline
		
		\multicolumn{5}{c}{With initial rankings generated by LambdaMart}\\
		DLCM & 0.6760 & 0.6810${}^+$ & 0.6990${}^+$ & 0.7430${}^+$ \\ 
		SetRank${}_{\mathrm{MSAB}}^{init}$ & \textbf{0.6837}${}^{+\dag}$ & 0.6820${}^+$ & 0.7009${}^+$ & 0.7443${}^+$ \\ 
		SetRank${}_{\mathrm{IMSAB}}^{init}$ & 0.6822${}^{+\dag}$ & \textbf{0.6835}${}^{+\dag}$ & 0.7029${}^+$ & 0.7453${}^{+\dag}$ \\ 
	    \hline
	\end{tabular}
\end{table}

\subsection{Experimental Results} \label{sec:results}

Table~\ref{Table.Experiments} lists the experimental results on the three datasets: Istella dataset, MSLR30K dataset, and Yahoo! dataset, respectively. For convenient comparisons, we categorized the baselines and the four versions of SetRank into three groups: models based on uni-variate scoring functions (i.e., RankSVM, RankBoost, Mart, and LambdaMart), models based on multivariate scoring functions and without initial rankings (i.e., DLCM${}^{w/o \, init}$, GSF, SetRank$_{MSAB}$, and SetRank$_{IMSAB}$), and models based on multivariate scoring functions and with initial rankings by LambdaMART (i.e., DLCM, SetRank$^{init}_{MSAB}$, and SetRank$^{init}_{IMSAB}$). 

Comparing the performances of the first group and the second (and the third) groups, we can see that the ranking models based on multivariate scoring functions are in general performed better than those based on uni-variate scoring functions. The multivariate models using LambdaMart as the initial rankings performed significantly better than LambdaMart, though LambdaMart is a very strong baseline. The results indicate that the multivariate scoring functions are superior to uni-variate ones because they captured the contextual information and document interactions during ranking. 

Also noted that SetRank outperformed the baselines of DLCM and GSF, in the cases of both with and without the initial rankings. In Istella dataset, without the initial rankings, SetRank${}_{\textrm{IMSAB}}$ improved DLCM${}^{w/o \, init}$ about 0.05 points in terms of NDCG@1; with the initial rankings, SetRank${init}_{\textrm{IMSAB}}$ improved DLCM about 0.02 points in term of NDCG@1. Similar phenomenon can be observed on other datasets with other evaluation measures. As discussed previously, the RNN in DLCM is very sensitive to the orders the document being inputted~\cite{ai2018learning}. SetRank, on the other hand, learns a permutation-invariant ranking model and thus is more robust. 

Comparing the performances of SetRank${}_{\textrm{MSAB}}$ and SetRank${}_{\textrm{IMSAB}}$, in most cases, SetRank${}_{\textrm{IMSAB}}$ achieved better document rankings. This is because the IMSAB blocks can automatically adapt to input set of any size, which makes the ranking model more robust.

\begin{figure}
    \centering
    \includegraphics[width=0.85\linewidth]{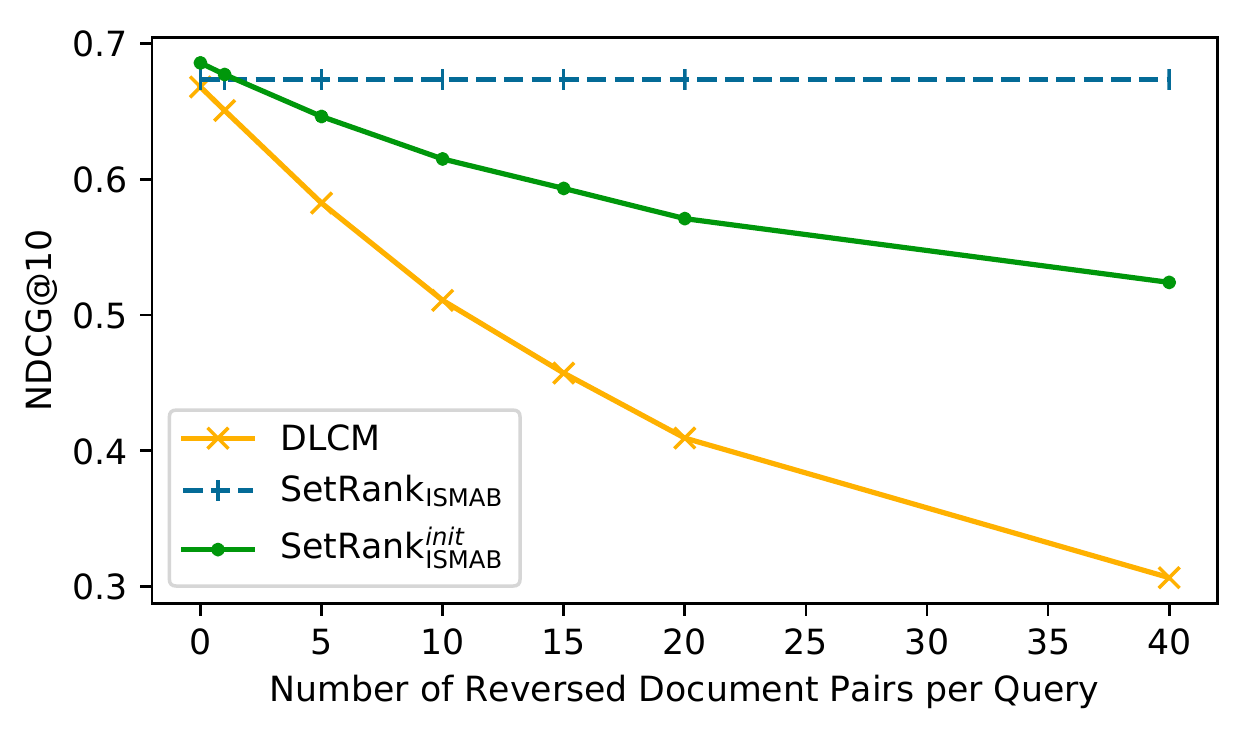}
    \caption{The performance of models w.r.t. the number reversed document pairs per query in the initial rankings.}
    \label{fig:perminv}
\end{figure}

\begin{table*}
    \centering
    \caption{Performances of the models trained and tested on queries that are associated with different number of documents. The values in the parentheses ($\Delta$) are the performance variations compared with the models trained and tested on set size 40.}
	\label{Table.Analyses1}
	\small
	\begin{tabular}{l l l l l l l}
		\hline
		 & \multicolumn{2}{c}{\# docs associated per query} & \multicolumn{4}{c}{} \\
		Model & training set ($N_0$) & test set ($N_1$) & NDCG@1 ($\Delta$) & NDCG@3 ($\Delta$) & NDCG@5 ($\Delta$) & NDCG@10 ($\Delta$) \\
		\hline
		\hline
		\multirow{5}{*}{DLCM${}^{w/o \, init}$} & \multirow{3}{*}{40} 
		       & 40  & 0.6233 & 0.5684 & 0.5825 & 0.6298 \\
		     & & 240 & 0.5943 (-0.0290) & 0.5394 (-0.0290) & 0.5518 (-0.0307) & 0.5964 (-0.0334) \\
		     & & 500 & 0.5844 (-0.0389) & 0.5300 (-0.0384) & 0.5428 (-0.0397) & 0.5868 (-0.0430) \\
		 \cline{2-7}
		 & 240 & 240 & 0.6199 & 0.5708 & 0.5836 & 0.6309 \\
		 & 500 & 500 & 0.6258 & 0.5700 & 0.5833 & 0.6298 \\ 
		\hline
		\multirow{5}{*}{SetRank${}_{\mathrm{MSAB}}$} & \multirow{3}{*}{40} 
		        & 40 & 0.6702 & 0.6150 & 0.6282 & 0.6766 \\
		     & & 240 & 0.6578 (-0.0124) & 0.6026 (-0.0124) & 0.6155 (-0.0127) & 0.6602 (-0.0164) \\
		     & & 500 & 0.6533 (-0.0169) & 0.5969 (-0.0181) & 0.6102 (-0.0180) & 0.6547 (-0.0219) \\
		 \cline{2-7}
		 & 240 & 240 & 0.6736 & 0.6141 & 0.6295 & 0.6777  \\
		 & 500 & 500 & 0.6712 & 0.6170 & 0.6316 & 0.6816 \\
		\hline
		\multirow{5}{*}{SetRank${}_{\mathrm{IMSAB}}$} & \multirow{3}{*}{40} 
		        & 40 & 0.6733 & 0.6136 & 0.6278 & 0.6737 \\
		     & & 240 & 0.6674 (-0.0059) & 0.6104 (-0.0032) & 0.6244 (-0.0034) & 0.6688 (-0.0049) \\
		     & & 500 & 0.6665 (-0.0068) & 0.6082 (-0.0054) & 0.6220 (-0.0058) & 0.6662 (-0.0075) \\
		 \cline{2-7}
		 & 240 & 240 & 0.6699 & 0.6115 & 0.6264 & 0.6750 \\
		 & 500 & 500 & 0.6696 & 0.6117 & 0.6247 & 0.6727 \\
		\hline
	\end{tabular}
\end{table*}

\subsection{Empirical Analyses} \label{sec:analyses}

We now investigate the reasons that SetRank outperforms the baseline methods, using the results on Istella dataset as examples. 

\subsubsection{The effects of permutation invariance}
We conducted experiments to show that permutation invariance property helps to learn a ranking model robust to the noises in the initial ranking list. Specifically, we randomly reversed some document pairs in the ranking lists generated by LambdaMart for each query. The processed document lists are then feeded as the initial rankings to DLCM and SetRank$_{\mathrm{ISMAB}}^{init}$ in the test phase, while both of DLCM and SetRank$_{\mathrm{ISMAB}}^{init}$ model are trained using the clean ranking list. 

Figure~\ref{fig:perminv} illustrates the performance curves in term of NDCG@10 w.r.t. the number reversed document pairs. We can see that, with the increasing number of reverse pairs (i.e., with more noise) in the initial rankings, both SetRank$^{init}_{\mathrm{ISMAB}}$ and DLCM performed worse. However, much more performance declines are observed with DLCM. This is because DLCM encodes the input documents with an RNN, which is sensitive to the permutations on the input set. 

We also illustrate the performance curve of SetRank${}_{\mathrm{ISMAB}}$, which did not use the ordinal embedding, in Figure~\ref{fig:perminv}. We can see that the performances of SetRank${}_{\mathrm{ISMAB}}$ did not change w.r.t. the number of reversed document pairs. The results confirmed the correctness of the theoretical analysis in Proposition~\ref{pro:PermuEqu}.

\subsubsection{The effects of set size adaptation}

In this part, we want to test the effects of ISMAB which enables the set size adaptation ability of SetRank.
We first trained and tested the DLCM and SetRank models (without the initial rankings) on the data that each query was associated with $N$ documents ($N=N_0=N_1=40, 240$, or 500). From the corresponding results shown in Table~\ref{Table.Analyses1} (lines in which ``\# docs associated per query'' in training set and in test set are identical), we can see that the three models, DLCM$^{w/o \, init}$, SetRank$_{MSAB}$ and SetRank$_{IMSAB}$ have similar performances on different set sizes. It indicates that when the document sets have similar sizes in phase of training and test, the performances of the models tend to be unchanged.
Nevertheless, when we trained the DLCM and SetRank models on queries associated with 40 documents ($N_0=40$) and tested the models on queries associated with 240 and 500 documents ($N_1=240$ or 500). From the corresponding results reported in Table~\ref{Table.Analyses1} (lines in which ``\# docs associated per query'' in training set is 40, and in test set is 240 or 500), we can see that the performance drop (see the $\Delta$ values in the brackets) in terms of NDCG@10 for DLCM$^{w/o \, init}$ is big, indicating that the sequential model DLCM$^{w/o \, init}$ is sensitive to the test set size. 
For example, when the models were trained on queries associated with 40 documents while tested on queries with 240 documents, the performance drop for DLCM$^{w/o \, init}$ is 0.0334, which is double compared to SetRank${}_{MSAB}$ and six times compared to SetRank${}_{IMSAB}$.

Additionally, note that the performance drop of SetRank${}_{MSAB}$ is much severe than that of for SetRank${}_{IMSAB}$, indicating that projecting variance set size to a fixed number of cluster centers as IMSAB does, reduces the sensitivity of the test set size. For example, when the models were trained on queries associated with 40 documents while tested on queries with 240 documents, the performance drop for SetRank${}_{IMSAB}$ in terms of NDCG@10 is 0.0049, while the number is 0.0164 for SetRank${}_{MSAB}$.

\subsubsection{The effects of initial rankings}

\begin{table}
    \centering
    \caption{Performances of SetRank variations that involve zero, one, or multiple initial rankings. }
	\label{Table.Analysis3}
	\small

	\begin{tabular}{l l l l l }
		\hline
		 & \multicolumn{4}{c}{NDCG} \\
		Model & @1 & @3 & @5 & @10 \\
		\hline
		SetRank${}_{\mathrm{IMSAB}}$ & 0.6733 & 0.6136 & 0.6278 & 0.6737 \\ 
		SetRank${}_{\mathrm{IMSAB}}^{init}$ & 0.6760 & 0.6202 & 0.6345 & 0.6834 \\ 
		\tabincell{l}{SetRank${}_{\mathrm{IMSAB}}^{multi-init}$} & 0.6744 & 0.6211 & 0.6365 & 0.6860  \\
		\addlinespace
		\hline
	\end{tabular}
\end{table}

One advantage of SetRank is that it can involve zero, one, or multiple initial document rankings as inputs, through adding one or more ordinal embeddings. We conducted experiments to test the effects of the initial rankings in SetRank. Specifically, we tested the performances of SetRank${}_{IMSAB}$ with no initial ranking, SetRank$^{init}_{IMSAB}$ with one initial ranking based on LambdaMart, and SetRank$^{multi-init}_{IMSAB}$ with four initial rankings based on LambdaMart, Mart, RankSVM, and RankBoost. Table~\ref{Table.Analysis3} reports the performances of these three SetRank variations in terms of NDCG at the positions of 1, 3, 5, and 10. From the results, we found that (1) SetRank$^{init}_{IMSAB}$ with one initial ranking performed better than SetRank${}_{IMSAB}$ which has no initial ranking, indicating that initial rankings do help to improve the performances; (2) SetRank$^{multi-init}_{IMSAB}$ with 4 initial rankings performed a little bit better than SetRank$^{init}_{IMSAB}$ which used only one initial ranking, indicating that  multiple rankings provide limited helps in improving the ranking accuracy. 
One possible reason is that LambdaMART has provided significantly better document rankings comparing to other ranking models, which makes the information obtained from other ranking models less useful in our experiments. 

\section{Conclusion} \label{sec:conclusion}
In this paper, we proposed a novel learning-to-rank model for information retrieval, referred to as SetRank. In contrast to existing models, the scoring function in SetRank is designed as a multivariate mapping from a document set to a permutation, and satisfy two critical requirements: cross-document interactions and permutation invariance. Self-attention mechanism in Transformer is employed to implement the scoring function. SetRank offers several advantages: efficiently capturing local context information, naturally involving (multiple) initial rankings, robust to input noise, and high accuracy in ranking. Experimental results on three large scale datasets show that SetRank outperformed the traditional learning-to-rank models and  state-of-the-art deep ranking models. Analyses showed that the two requirements did help to improve the performance and robustness of SetRank model. 

\begin{acks}
This work was supported by Beijing Academy of Artificial Intelligence (BAAI) under Grants BAAI2020ZJ0303, and BAAI2019ZD0305, the National Natural Science Foundation of China (NSFC) under Grants No. 61773362, 61872338, 61832017, and 61906180, the Youth Innovation Promotion Association CAS under Grants No. 2016102, the Tencent AI Lab Rhino-Bird Focused Research Program (No. JR202033), the National Key R\&D Program of China under Grants No. 2016QY02D0405, the Beijing Outstanding Young Scientist Program No. BJJWZYJH012019100020098.
\end{acks}



\newpage
\bibliographystyle{ACM-Reference-Format}
\bibliography{set2perm_sigir}

\appendix

\section{Proof to Proposition~\ref{pro:PermuEqu}}
\label{appendix:proof}
Note that the multi-head parameters $\mathbf{W}_i^Q, \mathbf{W}_i^K, \mathbf{W}_i^V$, the row-wise feedforward function $\textrm{rFF}(\cdot)$, and layer normalization $\textrm{LayerNorm}(\cdot)$ are all element wise functions, which do not affect the permutation-equivariant property. Therefore, we only need to consider the permutation-equivariant of the self-attention function:
\[
	\mathrm{SelfAttn}(\mathbf{X}) = \mathrm{Attn}(\mathbf{X}, \mathbf{X}, \mathbf{X}) = \mathrm{softmax}\left(\frac{\mathbf{XX}^T}{\sqrt{E}}\right)\mathbf{X}.
\]

\begin{proof}
	Let $\mathbf{X} = \{ \mathbf{x}_1, \mathbf{x}_2, \dots, \mathbf{x}_N \}$. For simplicity and with no loss of generality we discard the scalar factor $\sqrt{E}$, we have
	\[
	\begin{split}
		\mathrm{SelfAttn}(\mathbf{X}) &= \mathrm{softmax}\left(\mathbf{XX}^T\right)\mathbf{X} 
		= \mathrm{softmax}\left( \left[ \mathbf{x}_i \mathbf{x}_j^T \right]_{ij} \right)\mathbf{X} \\
		&= \left[ \frac{e^{\mathbf{x}_i \mathbf{x}_j}}{\sum_k e^{\mathbf{x}_i \mathbf{x}_k}} \right]_{ij}\mathbf{X} 
		=  \left[ \frac{\sum_j e^{\mathbf{x}_i \mathbf{x}_j} \cdot \mathbf{x}_j}{\sum_k e^{\mathbf{x}_i \mathbf{x}_k}} \right]_{i},
	\end{split}
	\]
	For any permutation $\pi \in \Pi_N$, since sum is permutation-invariant, 
	\[
	\begin{split}
		\mathrm{SelfAttn}(\pi \mathbf{X}) &= \mathrm{SelfAttn}\left( [\mathbf{x}_{\pi(i)}]_i \right)
		= \left[ \frac{\sum_j e^{\mathbf{x}_{\pi(i)} \mathbf{x}_{\pi(j)}} \cdot \mathbf{x}_{\pi(j)}}{\sum_k e^{\mathbf{x}_{\pi(i)} \mathbf{x}_{\pi(k)}}} \right]_{\pi(i)} \\
		&= \left[ \frac{\sum_j e^{\mathbf{x}_{\pi(i)} \mathbf{x}_{j}} \cdot \mathbf{x}_{j}}{\sum_k e^{\mathbf{x}_{\pi(i)} \mathbf{x}_{k}}} \right]_{\pi(i)}
		= \pi \left( \mathrm{SelfAttn}(\mathbf{X}) \right).
	\end{split}
	\] 
\end{proof}

\end{document}